\documentclass[sigconf]{acmart}
\AtBeginDocument{%
  }

\setcopyright{acmlicensed}
\copyrightyear{2018}
\acmYear{2018}
\acmDOI{XXXXXXX.XXXXXXX}
\acmConference[Conference acronym 'XX]{Make sure to enter the correct
  conference title from your rights confirmation email}{June 03--05,
  2018}{Woodstock, NY}
\acmISBN{978-1-4503-XXXX-X/2018/06}

\usepackage{amsmath}
\usepackage{algorithmic}
\usepackage{graphicx}
\usepackage{textcomp}
\usepackage{xcolor}
\usepackage{tikz}
\usetikzlibrary{shapes.geometric, arrows.meta, positioning}
\usepackage{pgfplots}
\pgfplotsset{compat=1.18}
\usepackage{xspace}
\usepackage{soul}
\sethlcolor{yellow!35}
\usepackage{multirow}
\usepackage{listings}
\usepackage[normalem]{ulem}

\usepackage{soul}
\soulregister\tool{0}

\newcommand{\citeme}[1]{%
  \begingroup
  \definecolor{hlcolor}{RGB}{255, 226, 176}\sethlcolor{hlcolor}%
  \textcolor{darkgray}{\hl{\textbf{CITE}}}%
  \endgroup
}

\newcommand{\tool}{\textsc{RepairFormer}\xspace}

\begin{document}

\title{\tool: Automated Repair of Structured Inputs Using Transformers}

\author{Ovi Paul}
\email{opaul@uh.edu}
\affiliation{%
  \institution{University of Houston}
  \city{Houston}
  \state{Texas}
  \country{USA}
}

\author{Tom J King}
\email{tjking2@uh.edu}
\affiliation{%
  \institution{University of Houston}
  \city{Houston}
  \state{Texas}
  \country{USA}
}

\author{Ali Shokri}
\email{ashokri@uh.edu}
\affiliation{%
  \institution{University of Houston}
  \city{Houston}
  \state{Texas}
  \country{USA}
}

\begin{abstract}
Structured input files such as JSON, DOT, OBJ, INI, S-expression, and TinyC are widely used in software systems, but small corruptions can cause parsers to reject otherwise useful data. Repairing such inputs is important because malformed configuration, program, and data files can interrupt testing, analysis, deployment, and downstream automation even when most of the original content remains intact. Existing repair techniques can produce structurally valid inputs, but they often rely on deletion or repeated search, which may lose original content and result in semantic incorrectness. This paper presents \tool, a transformer-based framework for structured input repair. The approach formulates repair as a supervised sequence generation task and uses format tags, oracle validation, and boundary-localized repair to generate valid outputs while preserving content. The boundary workflow focuses generation on the detected fault region, reducing the input size, and supporting repair of longer files. In evaluation, \tool achieves a 88\% in repair and 94\% in recovery, showing strongest content preservation when repairs are successful. Additional experiments on our benchmark shows \tool repairs 97.57\%  and recovers 94.29\%  with 5x faster runtime compared to state of the art.
\end{abstract}

\ccsdesc[500]{Software and its engineering~Software maintenance tools}

\maketitle

\section{Introduction}

Modern software systems rely on structured input formats such as JSON, XML, DOT, OBJ, INI, S-expression, and TinyC to represent configuration data, program structures, and domain specific information\cite{ddmax, erepair}. These files are often manually edited or generated by external tools, which makes them vulnerable to corruption. Even small syntax errors, such as missing delimiters, misplaced tokens, or incomplete structures, can cause a parser to reject the entire file\cite{korn}.
Such invalid inputs are common in real-world repositories (e.g., GitHub)~\cite{ddmax}, causing useful information to become inaccessible even when most of the file remains correct. Repairing such files manually can be slow and error-prone, especially for large files or when the error location is unclear.

Prior input repair techniques address this problem through parser guided search. For example, ddmax\cite{ddmax} repairs inputs by removing fragments until the remaining file is accepted by the parser. More recently, $\epsilon$REPAIR\cite{erepair} extends search-based repair by considering insertions, deletions, and replacements. But these approaches rely on many repeated oracle executions and iterative search steps to explore candidate repairs.

This work investigates a different direction. We formulate structured input repair as a sequence generation task using transformer-based model \cite{transformers, seqr}. Instead of only searching over edit operations, the model learns repair patterns from corrupted and valid input pairs. The goal is not only to produce a parser valid file, but also to preserve as much of the original information as possible. To improve scalability for long inputs, we also put a boundary localized workflow that guides the model toward the suspected faulty region before generation. RepairFormer can support the recovery of malformed configuration and
other structured input files before downstream processing
\cite{config,korn,ddmax}.
This paper makes the following contributions:
\begin{itemize}

\item We present \tool, a transformer-based tool that repairs structured inputs corrupted through a variety of transformations, including deletions, insertions, and byte flips, while maximizing content preservation and reducing the cost of search-based repair.
\item We analyze format-wise behavior, mutation types, runtime, and model choices to identify when learned repair is most effective.
\end{itemize}

\textbf{Tool Availability:}
The source code of \tool is publicly available through its GitHub repository~\footnote{https://github.com/pass-uh/RepairFormer.git}.

\begin{figure*}[t]
    \centering
    \includegraphics[width=0.8\linewidth]{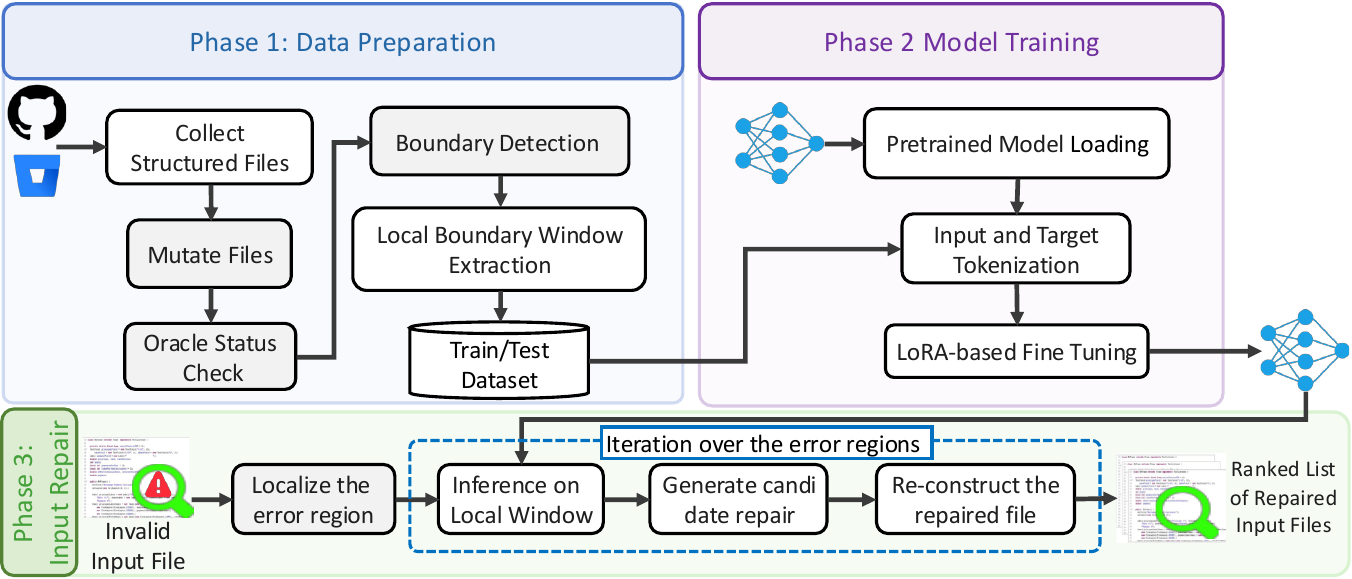}
    \caption{Overview of the \tool}
    \label{fig:overall}
\end{figure*}

\section{Motivating Example}
Consider the following corrupted JSON input:
\begin{quote}
\texttt{\{"id": 17, "status" "active"\}}
\end{quote}
The input is invalid because the colon after \texttt{"status"} is missing. A deletion-based repair method may produce a valid file by removing the corrupted field:
\begin{quote}
\texttt{\{"id": 17\}}
\end{quote}
This output is syntactically valid, but it loses the status information. While the state of the art tools either produce the above output or randomly generate characters to repair the file, a generative repair model can instead preserve the field by inserting the missing colon:
\begin{quote}
\texttt{\{"id": 17, "status": "active"\}}
\end{quote}
This example shows the importance of preserving the useful content by an input repair tool while also satisfying the syntax correctness of a structured input.

\section{An Overview of \tool}
\label{sec:approach}
\tool treats input repair as a supervised sequence generation task in which, given an invalid structured input, the tool generates a repaired version that satisfies both the format (i.e., syntactic) and the intended structural and semantic constraints of the input. Figure \ref{fig:overall} provides an overview of the tool. We first create a dataset of training data from $<buggy, repaired>$ input pairs (Section \ref{sec:dataset_construction}), followed by training the model (Section \ref{sec:model_training}). To better pinpoint the buggy part of the input during the repair process, \tool localizes that part within the input file and fixes the issue (Section \ref{sec:input_repair}).

\subsection{Dataset Preparation}
\label{sec:dataset_construction}

\begin{table}[t]
\centering
\small
\caption{Valid structured data collected.}
\label{tab:dataset_stats}
\begin{tabular}{lrr}
\toprule
Format & Files & Size range \\
\midrule
DOT   & 1154 & 23--19856 bytes \\
INI   & 1054 & 9--19835 bytes \\
JSON  & 901  & 9--19647 bytes \\
OBJ   & 888  & 38--19994 bytes \\
S-expression & 773  & 10--16281 bytes \\
TinyC & 1000 & 2--636 bytes \\
\bottomrule
\end{tabular}
\end{table}

The dataset is constructed from valid structured files collected from GitHub. We used a script that searches GitHub by file extension and size range, downloads candidate files, removes duplicates using SHA256 hashes, and validates each file with an oracle. Invalid samples are automatically synthesized through controlled mutations, including single character corruption, double character corruption, and truncation. Table \ref{tab:dataset_stats} provides distribution information of the created dataset.
A format tag, such as \texttt{json}, \texttt{ini}, \texttt{dot}, \texttt{obj}, \texttt{s-expression}, or \texttt{tinyc}, is prepended to each input so that the model can support multiple formats.
Each mutated file is verified by the corresponding oracle to ensure invalidity, while the original file is kept as the repair target. The mutation strategy was adopted from $\epsilon$REPAIR\cite{erepair}, which was extended from the approach introduced in DDMax\cite{ddmax}.
Each valid and invalid pair is converted into a JSONL record:
$\{x = \texttt{format: invalid\_text}, \quad y = \texttt{valid\_text}\}$, 
where \(x\) is the model input and \(y\) is the target repair. The conversion script loads each mutation pair, assigns it to the proper train or test split, and writes train, validation, and test JSONL files. The validation set is formed by randomly selecting 10\% of the training data.

\subsection{Model Training}
\label{sec:model_training}
The model is initialized from \texttt{Salesforce/codet5-base} \cite{codet5}, which is a variant of T5 model \cite{t5}. This variant was trained on large corpus of program data, so this is suitable for working with structured input data. The input and target fields are loaded from JSONL files, tokenized with the CodeT5 tokenizer, and passed to a sequence generation trainer. The model is trained to minimize validation loss, and early stopping is used to prevent unnecessary training after convergence. The training configuration supports model fine tuning and low rank adaptation(LoRA)-based~\cite{lora} parameter efficient fine tuning. When LoRA is enabled, low rank adapters are applied to the attention projection modules, which reduces the number of trainable parameters while keeping the base model fixed.

To narrow down the repair process to the actual buggy component, we utilize the boundary localization module from $\epsilon$REPAIR~\cite{erepair}. To that end, instead of passing the entire invalid file to the model, \tool identifies a suspected error boundary and extracts a smaller, localized window around it.

The boundary can be estimated using oracle-based boundary detection, mutation metadata, or the first-byte difference between the valid and invalid files. The oracle-based localization follows the boundary search paradigm introduced in $\epsilon$REPAIR~\cite{erepair}, where parser feedback is leveraged to distinguish among correct, incomplete, and incorrect prefixes. A binary search is then executed to locate the largest prefix that is not explicitly evaluated as incorrect, and this index is designated as the repair boundary.

After localization, a local context window surrounding the boundary is extracted. Let $s_i$ denote the original valid file string and $\tilde{s}_i$ represent its mutated, invalid counterpart. We define the slicing notation $s_{i,a:b}$ as the substring of $s_i$ from byte index $a$ to $b$. Here, $x_i$ represent the localized input extracted from the $i$th invalid file. The extracted window from the invalid file is marked with a \texttt{<BOUNDARY>} token at the pinpointed fault location:
\[
x_i = \text{\texttt{format: }} 
\tilde{s}_{i,a:b}^{\,\text{\texttt{BOUNDARY}}}.
\]

Similarly, $y_i$ represent the corresponding target sequence extracted from the original valid file. The target for sequence generation is the corresponding sequence window from the original valid file:
\[
y_i = s_{i,a':b'},
\]
where $a'$ and $b'$ represent the adjusted alignment boundaries in the clean document. The window size is selected under the model token budget so that the localized prompt remains within the context limit.

This workflow has two main benefits. First, it reduces the effective input length for long files. Second, it makes the repair task more explicit by directing the model to the suspected faulty region. 

\subsection{Input Repair}
\label{sec:input_repair}

During the repair process, \tool takes an input file and first invokes the parser as an oracle to determine whether the file is invalid. If the file is deemed invalid, the tool locates the suspected error region, extracts a local window around that region, and inserts the \texttt{<BOUNDARY>} marker into the input before passing it to the repair model. The trained model then generates a repair candidate for that local window. Model inference refers to the generation of this repair candidate, rather than the localization or extraction of the local window. The generated candidate is reconstructed into the original file and validated again using the oracle. A repair is accepted when the reconstructed file becomes valid. Invalid candidates are rejected unless they advance the error boundary. The process repeats until the file is repaired or the maximum number of iterations is reached.

\section{Evaluation}
We evaluate \tool on invalid structured inputs spanning multiple formats and mutation types. The evaluation focuses on the tool's ability to transform invalid inputs into parser-accepted valid files while preserving the original content of unaffected regions. As discussed later, we compare \tool against state-of-the-art repair tools and report both format- and mutation-specific results. We further analyze runtime performance and the number of repair iterations required to characterize the computational cost of the repair process.

\subsection{Evaluation Metrics}
In our experiments, the repaired output is evaluated using three primary metrics: \textit{repair}, \textit{recovery}, and \textit{runtime}. A file is counted as \textit{repaired} when the original input is invalid and the selected model output becomes valid under the oracle. Repair measures the percentage of invalid files that are successfully transformed into oracle-valid outputs:
$\text{Repaired} =
\frac{\text{Valid Input Files}}{\text{Originaly Invalid Input Files}} \times 100.
$
Data preservation is measured using normalized Levenshtein distance \cite{levenshtein} between the repaired output and the original valid target:
$
\text{DataLoss}(s,\hat{s}) =
\frac{d_{\text{lev}}(s,\hat{s})}{\max(1, |s|)} \times 100.
$
Here, \(s\) is the original valid file, \(\hat{s}\) is the repaired output, and \(d_{\text{lev}}\) is the Levenshtein edit distance. Lower data loss means the repaired output preserves more of the original file content. \textit{Recovery} is reported as the complement of data loss:
$
\text{Recovered}(s,\hat{s}) =
100 - \text{DataLoss}(s,\hat{s}).
$
Finally, \textit{runtime} measures the computational cost of repair. We report the total time required to process a file, including boundary localization, repair generation, candidate validation, oracle checking, splicing, and output writing. To identify the primary bottlenecks, we also report generation and oracle-validation times separately.

\subsection{Results}

\begin{table}[t]
\centering
\caption{Performance comparison across two benchmarks.}
\label{tab:combined_repair_results}
\renewcommand{\arraystretch}{1.15}
\setlength{\tabcolsep}{2pt}
\small
\begin{tabular}{@{}lcccccccc@{}}
\toprule
\multirow{2}{*}{\textbf{Method}} 
& \multicolumn{4}{c}{\textbf{$\epsilon$Repair Benchmark}} 
& \multicolumn{4}{c}{\textbf{\tool Benchmark}} \\
\cmidrule(r){2-5} \cmidrule(l){6-9}
& \textbf{Rep.} & \textbf{Rec.} & \textbf{Tot.} & \textbf{Time}
& \textbf{Rep.} & \textbf{Rec.} & \textbf{Tot.} & \textbf{Time} \\
\midrule

$\epsilon$REPAIR~\cite{erepair}
& 97\% & 92\% & \textbf{89.24\%} & 3.87s
& 87.91\% & 94.03\% & 82.66\% & 48.55s \\

ANTLR~\cite{antlr}
& 49\% & 90\% & 44.1\% & \textbf{0.31s}
& 27.14\% & 89.23\% & 24.22\% & 12.43s \\

DDMax~\cite{ddmax}
& \textbf{98\%} & 81\% & 79.38\% & 2.71s
& 36.18\% & 86.27\% & 31.21\% & 137.20s \\

\tool
& 88\% & \textbf{94\%} & 82.92\% & 7.16s
& \textbf{97.57\%} & \textbf{94.29\%} & \textbf{92\%} & \textbf{9.71s} \\

\bottomrule
\end{tabular}
\end{table}

Table \ref{tab:combined_repair_results} compares \tool with $\epsilon$REPAIR, ANTLR, and DDMax on two benchmarks, with bold values indicating the best scores. On the external $\epsilon$REPAIR benchmark, \tool achieves 88\% repair, 94\% recovery, and an overall score of 82.92\%. It outperforms ANTLR in repair and recovery. Although DDMax achieves higher repair at 98\%, \tool improves recovery from 81\% to 94\% and the overall score by 3.54\%, showing that it preserves more content than deletion based repair.

The higher performance on the \tool benchmark likely reflects closer alignment with its training distribution. The $\epsilon$REPAIR benchmark contains different Lisp structures and mostly single mutations, while the \tool benchmark contains more double mutations and truncations. This domain shift makes $\epsilon$REPAIR an important external generalization test.

On the \tool benchmark, \tool improves repair over $\epsilon$REPAIR from 87.91\% to 97.57\% and recovery from 94.03\% to 94.29\%, while reducing runtime from 48.55 to 9.71 seconds, approximately 5$\times$ faster. The longer runtimes of $\epsilon$REPAIR and DDMax are caused by failed cases reaching the four minute timeout. The difference is influenced by the more balanced mutation distribution of the \tool benchmark compared with the predominantly single mutations in $\epsilon$REPAIR. Overall, \tool is more effective and efficient on larger inputs.

\begin{table}[t]
\centering
\caption{Format and mutation wise metrics.}
\label{tab:mutation_type_analysis}
\renewcommand{\arraystretch}{1.15}
\setlength{\tabcolsep}{4pt}
\small
\begin{tabular}{llccc}
\hline
\textbf{Format} & \textbf{Case} & \textbf{Repair} & \textbf{Recovery} & \textbf{Runtime} \\
\hline
C    & Single C      & 97.50\% & 95.18\% & 0.98 s \\
C    & Double C      & 92.00\% & 91.47\% & 1.63 s \\
C    & Truncated C   & 99.00\% & 76.78\% & 0.81 s \\
\hline
DOT  & Single DOT    & 95.30\% & 99.69\% & 10.47 s \\
DOT  & Double DOT    & 96.00\% & 99.70\% & 15.69 s \\
\hline
INI  & Single INI    & 93.40\% & 99.48\% & 11.13 s \\
INI  & Double INI    & 49.00\% & 99.81\% & 10.17 s \\
\hline
JSON & Single JSON   & 94.44\% & 99.65\% & 11.61 s \\
JSON & Double JSON   & 91.92\% & 99.46\% & 18.69 s \\
\hline
S-EXPRESSION  & Single S-EXPRESSION    & 61.70\% & 80.99\% & 1.97 s \\
S-EXPRESSION  & Double S-EXPRESSION   & 65.00\% & 60.60\% & 2.31 s \\
\hline
OBJ  & Single OBJ    & 94.60\% & 99.53\% & 12.38 s \\
OBJ  & Double OBJ    & 88.00\% & 99.57\% & 16.81 s \\
OBJ  & Truncated OBJ & 79.00\% & 91.22\% & 14.56 s \\
\hline
\end{tabular}
\end{table}

Table~\ref{tab:mutation_type_analysis} shows that performance varies by format and mutation type. DOT and C achieve repair above 92\% for single and double mutations, while successful repairs for INI, JSON, and OBJ generally recover more than 91\% of the original content.

Double mutations generally increase runtime because they require more complex corrections or additional attempts, while truncation reduces recovery because missing content may not be reconstructed. S expression is the most difficult format, particularly under double mutation. The performance decreases mainly for truncated inputs and highly nested S expression files.

\begin{figure}
    \centering
    \includegraphics[width=0.95\linewidth]{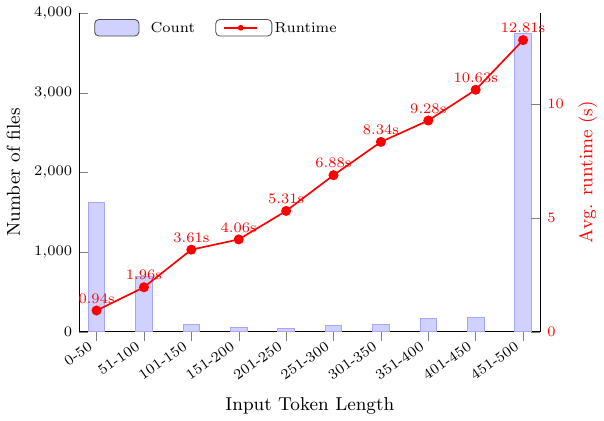}
    \caption{Input token length distribution with average runtime. Bars show the number of files in each token range, and the red line shows the average total runtime in seconds.}
    \label{fig:token_runtime_histogram}
\end{figure}

Figure~\ref{fig:token_runtime_histogram} shows that runtime increases with input token length rather than file size. Short localized windows use fewer tokens and run faster, while inputs near the token limit require approximately 9.71 seconds. Reducing the token budget may lower runtime but can remove useful repair context.

\section{Related Work}
Several techniques have been proposed for debugging and repairing invalid program inputs. These approaches can be broadly categorized as parser-based recovery methods and automated input repair techniques.

Parser generators such as ANTLR~\cite{antlr} provide built-in error recovery for malformed inputs \cite{erepair}. However, they require a formal grammar specification and are typically limited to local syntax errors, making them less suitable for complex structural corruptions or formats without readily available grammars.

DDMax~\cite{ddmax} repairs corrupted structured inputs by identifying the largest parser-accepted subset of the input. While it is grammar-independent and supports multiple formats, it primarily repairs inputs through deletion, which can reduce content preservation. Its reliance on repeated parser executions can also increase runtime.

$\epsilon$REPAIR~\cite{erepair} is a format-independent repair technique that uses parser feedback to search for edits that transform invalid inputs into valid ones. It supports insertions, deletions, and replacements without requiring a grammar specification, but still relies on iterative search and repeated oracle executions. In contrast, \tool uses localized error context and a learned repair model to generate repairs directly.

\section{Limitations}
While \tool can effectively repair invalid input files, it has several limitations. First, the model is constrained by a maximum token length. Therefore, longer files may require more precise boundary localization or larger models. Second, repair outcomes depend on oracle-based validation, meaning that different parsers may accept different repairs for the same input. Third, if an error occurs near the edge of the extracted local context window, the model may fail to generate a correct repair. Finally, local repairs may be insufficient when producing a valid fix requires information from distant parts of the input.

\section{Conclusion}
This paper presented \tool, a transformer-based framework for repairing corrupted structured inputs. The approach formulates input repair as a sequence generation task and uses format tags, oracle validation, and boundary-localized repair to generate valid outputs while preserving original content. Compared with deletion-based repair methods, the proposed method focuses on reconstructing missing or corrupted syntax instead of removing large input fragments.

The evaluation shows that our approach achieves 88\% repair and 94\% recovery, indicating strong content preservation. It performs well on C, DOT, INI, JSON, and OBJ, while S expression remains more challenging. On our benchmark, \tool improves repair by about 10\% over state of the art methods while running 5x faster.

Overall, the results suggest that transformer-based input repair is a promising direction for producing content-preserving repairs across multiple structured formats.

\begin{acks}
ChatGPT~\cite{achiam2023gpt} was used to assist with the generation of limited text and code during the preparation of this work. All generated content was reviewed, validated, and revised by the authors. The authors are solely responsible for the research ideas, methodology, implementation, experiments, analysis, and conclusions presented in this paper. 

\end{acks}

\bibliographystyle{ACM-Reference-Format}
\bibliography{bibliography}

\end{document}